\documentclass[aps,prl,preprint]{revtex4}

\usepackage{epsfig}

\begin{document}
\title{The meson $B_c$ annihilation to leptons and inclusive light hadrons}

\author{Xing-Gang Wu\(^{1}\)\footnote{email: xgwu@itp.ac.cn},
Chao-Hsi Chang\(^{1,2}\) \footnote{email: zhangzx@itp.ac.cn},
Yu-Qi Chen\(^{1,2}\)\footnote{email: ychen@itp.ac.cn} and
Zheng-Yun Fang\(^{3}\)\footnote{email: zyfang@cqu.edu.cn}}

\address{\(^1\)Institute of Theoretical Physics, Chinese Academy of Sciences,
P.O.Box 2735, Beijing 100080, China}
\address{\(^2\)CCAST (World Laboratory), P.O.Box 8730, Beijing 100080, China}
\address{\(^3\)Department of Physics, Chongqing University,
Chongqing 400044, China}


\begin{abstract}
The annihilation of the $B_c$ meson to leptons and inclusive
light hadrons is analyzed in the framework of nonrelativistic QCD
(NRQCD) factorization. We find that the decay mode, which escapes from
the helicity suppression, contributes a sizable fraction width. According to
the analysis, the branching ratio due to the contribution from the color-singlet
component of the meson $B_c$ can be of order \(10^{-2}\). We also estimate the
contributions from the color-octet components. With the velocity scaling
rule of NRQCD, we find that the color-octet contributions are sizable too,
especially, in certain phase space of the annihilation they are greater
than (or comparative to) the color-singlet component. A few observables
relevant to the spectrum of charged lepton are suggested, that may be used as
measurements on the color-octet and color-singlet components in the future
$B_c$ experiments. A typical long distance contribution in the annihilation
is estimated too.\\


\end{abstract}

\maketitle

\section{I. Introduction}

The observation of the $B_c$ meson by CDF collaboration at
Fermilab\cite{CDF} is one of the important discoveries in heavy quark
physics and their results are consistent of the theoretical
predictions\cite{prod,chen,CCWZ,dec,MG,ZL}, when the theoretical
uncertainties and experimental errors are concerned. More $B_c$
events in the new run of the Tevatron than those of the discovery
and much more events by several orders at the hadron collider LHC
are expected\cite{prod}. Therefore, thorough experimental study of the
$B_c$ meson and more precise comparisons with theoretical predictions
on its properties, especially, on its various decay modes will be available
in foreseen future.

One of the interesting decay modes is its pure leptonic decay.
Since the decay amplitude is proportional to the decay constant
$f_{B_c}$ and the CKM matrix element $V_{cb}$. In principle, it
can be used to measure $f_{B_c}$ if we know the value of
$V_{cb}$, or vice versa. However, the pure leptonic decay modes
suffer by the the helicity suppression with small leptonic
mass. The suppression is not severe for $ B_c \to \tau \nu_\tau $
decay mode only, whereas, this decay mode is hard to be detected
because its resultant state contains two neutrinos at least.

There is no helicity suppression if more particles are included in
the decays. Such an example is the radiative leptonic decay,
\(B_{c} \rightarrow \gamma l^{+} \nu_{l} \), which has been
analyzed in \cite{chang,CCL,chang1,bct,TMA}. The results show that the
decay branching ratios are order of \(10^{-5}\) for  \( l =(\mu,
e)\), which is much larger than that of the pure leptonic decay of
$e$-channel and comparable to that of $\mu$-channel. The relatively larger
branching ratio may make them possibly to be measured at the LHC.

In this paper, we analyze a new annihilation decay mode of the meson
$B_c$, i.e., the `semi-inclusive' decay $B_c \rightarrow  l^{+} \nu_{l}
\, + $ {\it light hadrons}. Here the light hadrons are produced from
the emitted soft or/and hard gluons, i.e., we are not interested in any
specific light hadron(s) but sum all of them.  The annihilation
process escapes from the the helicity suppression due to additional
gluons in the final state. Since it is an annihilation of a pair heavy quarks
essentially, so it can fully be analyzed in the framework of Nonrelativisitic
QCD (NRQCD) factorization\cite{GEP}. According to the framework, the decay
width can be factored into a sum of the products of short distance coefficients and
NRQCD long distance matrix elements. The short distance coefficients can
be expanded as power series of $\alpha_s$ at the energy scale of
the heavy quark masses, thus they are calculable perturbatively.
We may attract them by matching the results obtained with QCD full theory
calculations at the threshold to the ones obtained by NRQCD effective theory.
Generally, the long distance matrix elements can be estimated only
by means of the velocity power counting rule of the NRQCD effective theory.

The effective theory NRQCD considers a quarkonium
state as an expansion of various Fock states.
The annihilation may be via various Fock states and
will depend on various long distance matrix elements.
According to NRQCD, the annihilation
\(B_{c} \rightarrow l^{+} \nu_{l} \)+ {\it light hadrons}
can be carried out via leading Fock states $|c\bar b_1(^{1}S_{0})> $,
$|c\bar b_8(^{3}S_{1}) g> $ and $|c\bar b_8(^{1}P_{1}) g> $ etc.
With a naive order estimate, for the leading order the concerned annihilation
may be the processes via the leading Fock state in velocity $v$ but an order higher
in QCD, i.e., $ (c\bar b)_1(^{1}S_{0}) \rightarrow l^{+} \nu_{l} + g g $, and those
via a higher Fock state but an order lower in QCD comparatively, such as
$(c\bar b)_8(^{3}S_{1})\rightarrow l^{+} \nu_{l} + g $ and $(c\bar b)_8(^1P_{1})
\rightarrow l^{+} \nu_{l} + g $ etc, because of the compensation
for the Fock states and QCD couplings. Therefore, in the decay
\(B_{c} \rightarrow l^{+} \nu_{l} \)+ {\it light hadrons} we should consider
the various short distance processes. Potentially the annihilation can be used
to probe the contributions from various components of the higher Fock states,
and it is why we investigate the annihilation quantitatively in this paper in
the framework of NRQCD factorization. Namely we estimate the decay rates accordingly
by taking the potential model value for the color-singlet matrix element which relates
to the wave function (derivative of the wave function) at origin directly and by means
of the velocity scaling rule of NRQCD for the color-octet ones.

In conventional potential model, a heavy quarkonium
just is a bound state of the heavy quark and antiquark
in color singlet\cite{ck}. Thus in the framework of potential model,
when the annihilation of the meson $B_c$ is estimated, only the
color-singlet component is taken into account. As a result, only
color-singlet matrix elements in NRQCD may be related to the wave functions
obtained in potential model.

In Ref.\cite{CCL}, some typical long distance contributions to the leptonic
radiative decay have been taken into account, and results show that such
contributions are negligible in comparison with the short distance
contributions. In fact in the annihilation concerned here, there are
also similar long-distance contributions. Therefore in this paper we
also estimate the long distance contributions.

The paper is organized as follows: The first section contributes to
the introduction. In section II, we briefly outline the formula for the
annihilation within NRQCD framework first, then in subsection $A.$ we
present the method to calculate the short distance coefficients for
color-singlet one and final result, which corresponds to the process
\((c\bar{b})_1\rightarrow l^{+} \nu_{l} gg \); and in subsection $B.$,
we calculate the short distance coefficients for color-octet ones, which
relates to \((c\bar{b})_8 \rightarrow l^{+} \nu_{l} g \) of higher Fock states
(color octet ones). In section III, we estimate the typical long distance
contributions of the color singlet leptonic decays using the so-called
`generalized instantaneous approximation'\cite{chen,CCWZ}.
In section IV, we present the numerical results, a brief summary and discussions.


\section{II. NRQCD factorization analysis}

In the non-relativistic heavy quark limit, there are several distinct energy
scales in heavy quark and antiquark bound state system. The heavy quark masses
$m_c$ and $m_b$ relate to the largest energy scale, while the off-shell 3-momentum
$mv$ and the typical binding energy $mv^2$ are the small ones, where $v\sim O(\alpha_s)$
is the relative velocity between the heavy quark and the antiquark. NRQCD effective
theory is established by integrating out the energy scale effects at heavy quark mass.
In the effective theory, the heavy quark and antiquark are described by non-relativistic
two component fields.
A physical state of $B_c$ meson can be decomposed into a sum of a set of Fock states:
$$|B_c\rangle=O(v^{0})|c\bar{b}_1(^{1}S_{0})\rangle
+O(v^1)|c\bar{b}_8(^{1}P_{1})g\rangle
+O(v^{1})|c\bar{b}_8(^{3}S_{1})g\rangle
+...\;.$$
According to the velocity scaling rule as indicated in the above expension,
the probability of each Fock state scales as a definite power
of $v$. Namely the leading Fock state of the
$B_c$ is $|c\bar{b}_1(^1S_0)\rangle$, whose probability is of
order $O(v^0)$. The next leading Fock states are $|c\bar{b}_8(^1P_1)g\rangle$
and $|c\bar{b}_8(^3S_1)g \rangle$, whose probability is of order
$v^2$. In fact, `certain order' in $v$ comes from
the integration for phase space, but here we restrict ourselves to consider the
wave function expansion itself so we do not take into account those
from relevant phase space integration. Throughout this paper we denote a
state of $(c\bar{b})$ pair with a subscript for color: $1$ for color singlet and $8$
for color-octet, but for its angular momentum quantum number we put in the
parentheses accordingly.

The annihilation of the $c$ and the $\bar{b}$ quarks happens in short-distance
comparing with perturbative QCD (pQCD) energy scale, because of $c$ and $\bar{b}$
being heavy. The characteristic space-time of the annihilation is of order $O(1/m_c)$
or $O(1/m_b)$. The size of the $B_c$ meson is of order $O({\frac{1}{v\mu}})$, where
$\mu=m_b\, m_c/(m_b+ m_c)$ is the reduce mass of the $\bar{b}$ and $c$
system. They are distinctly separated before annihilating. Thus NRQCD factorization
formula can also be used to analyze the decay of the $B_c$ meson.
We adopt the formulism of NRQCD to do calculations in this paper without
strong modification. In the annihilation $B_c\rightarrow l^{+} \nu_{l} \, + $
{\it light hadrons }, the
$\bar{b}$, $c$ quarks annihilate by weak and strong interactions both,
that, according to pQCD, is of the short-distance
effects, because it happens in such a short space-time at an order
$\ll 1/\lambda_{QCD}$. According to the NRQCD
factorization formula, the decay width can be factored into
\begin{eqnarray}\label{nrqcd}
  \Gamma &=& {1 \over 2 M }\sum_n { C_n  }
  \langle B_c |\hat{\it O}_n| B_c \rangle \;,
\end{eqnarray}
where $C_n (n=1,2,\cdots)$ are the short distance coefficients, and \( \langle
B_c |\hat{\it O}_n| B_c \rangle (n=1,2,\cdots) \) are NRQCD long distance matrix
elements. The short distance coefficients, being expanded in power
of $\alpha_s$, can be calculated with pQCD. While the long distance
matrix elements are the `average' values of the operators $\hat{\it O}_n,
n=1,2, \cdots$ which, being local gauge-invariant and relevant to the annihilation,
consists of non-relativistic 4-quark operators.
They measure the inclusive probability of finding a pair of $c\bar{b}$
being at the same point and with suitable color and angular-momentums
being specified.  By the {\it the velocity scaling rule} the possibility in the
meson $B_c$ in order of $v$.
The $(\bar{b} c)$ pair in the meson $B_c$, not only
in color-singlet but also in color-octet, may be annihilated at short-distance.
Correspondingly, the matrix element is regarded as a color-single
one or a color-octet one. The lowest order process for the color-singlet,
i.e. the short-distance annihilation of the meson $B_c$ to leptons and
light-hadrons is $(\bar{b}c)_1 \to l\nu_l + gg$, while those for
the color-octet ones are $(\bar{b}c)_8(^1P_1) \to l\nu_l + g$ and
$(\bar{b}c)_8(^3S_1) \to l\nu_l + g$. Hence the factorization reads
as the following form:
\begin{eqnarray}
\Gamma &=& {1\over 2 M} \Big[\;
 C_1(^1S_0)\;
  \frac{1}  {M^2}\;
  \langle B_c |
  \psi_c^\dag  \chi_b\; \chi_b^\dag  \psi_c
  | B_c \rangle
  \nonumber \\
  &+&\;
   C_8(^3S_1)\;
    \frac{1} {M^2}\;
  \langle B_c |
  \psi_c^\dag \sigma^i T^a \chi_b\; \chi_b^\dag \sigma^i T^a \psi_c
  | B_c \rangle
  \nonumber \\
  &+&\;
  C_8(^1P_1)\;
    \frac{1} {M^4}\;
  \langle B_c |
  \psi_c^\dag (-i\frac{1}{2}\stackrel
{\leftrightarrow}{D})T^a \chi_b\; \chi_b^\dag
(-i\frac{1}{2}\stackrel {\leftrightarrow}{D}) T^a \psi_c
  | B_c \rangle \;\Big] \;,
  \label{fact}
\end{eqnarray}
where $C_1(^1S_0)$, $C_8(^3S_1)$ and $C_8(^1P_1)$ are
dimensionless short distance coefficients and they are
of order $G_F \alpha_s^2$, $G_F \alpha_s$, and
$G_F \alpha_s$ (the lowest order in weak interaction with suitable
strong couplings) respectively. The matrix elements of the three
term in the bracket scales as $v^3$, $v^5$ and $v^5$,
respectively. Note that here we adopt
the normalization for the meson states as $\langle B_c({\bf
P'})|B_c ({\bf P})\rangle=2E_P(2\pi)^3\delta^3({\bf P'}-{\bf
P})$ and since we focus only those annihilations,
in addition to the leptons, also to light hadrons,
i.e. the pure leptonic annihilation of the
meson $B_c$ is not included, hence the contributions
from pure leptonic annihilation should not be included in the
coefficient $C_1(^1S_0)$ appearing in Eq.(\ref{fact}).

Now let us to compute these short-distance coefficients. Namely
we calculate the processes with pQCD precisely, and by matching the results
with the NRQCD factorized formula, finally we will obtain the coefficients.

The matrix element for the decay $B_c(P) \to l^+(k_4) + \nu_l(k_5)
+ X$ can generally be written as
\begin{equation}
  M=\frac{G_{F}V_{cb}}{\sqrt{2}} \left \langle l\nu_{l} | j^{\mu} | 0
    \right \rangle \langle X | J_{\mu} | B_{c} \rangle \;,
    \label{eq:matrix}
\end{equation}
where \(V_{cb}\) is the CKM matrix element,
and $j^\mu$ and $J_\mu$ are the weak currents for $l\nu$ leptons
and $c\bar{b}$ quarks.

The decay width can then be expressed as
\begin{eqnarray} \label{eq:inwidth}
&&\Gamma = \frac{G_F^2V_{cb}^2}{4M(2\pi)^6}\int
\frac{d^3\vec{k}_4}{{2 \varepsilon_{4}}}\frac{d^3\vec{k}_5}{{2
\varepsilon_{5}}} L^{\mu\nu}(k_4,k_5) \cdot t_{\mu\nu}(P,k_4+k_5)
\;,
\end{eqnarray}
where
\begin{eqnarray}
L^{\mu\nu}(k_4,k_5) &=& \bar{u}(k_5)\gamma^\mu(1-\gamma_5)v(k_4)
\bar{v}(k_4)\gamma^\nu(1-\gamma_5)u(k_5)\;, \nonumber \\
t_{\mu\nu}(P,k) &=& {\rm Im} \int d^4x e^{ikx} \langle
B_c(P)|J_\mu^w(x)J_\nu^w(0)|B_c(P)\rangle \;,\label{l-t}
\end{eqnarray}
are the leptonic tensor and the hadronic one, respectively.
The  $c\bar{b}$ quark currents in the hadronic tensor precisely are
$J_\mu^w=\bar{\psi_c}\gamma_\mu(1-\gamma_5)\psi_b$
and $J_\nu^w=\bar{\psi_b}\gamma_\nu(1-\gamma_5)\psi_c$.

The leptonic tensor $L^{\mu \nu}$ can easily be calculated and
reads:
\begin{equation}
L^{\mu \nu}=8 (k^{\mu}_{4} k^{\nu}_{5}+k^{\mu}_{5}
k^{\nu}_{4}-g^{\mu \nu}(k_{4}\cdot k_{5})-i \varepsilon^{\mu \nu
\alpha \beta} k_{4 \alpha} k_{5 \beta}),
\end{equation}
where \(\varepsilon^{\mu \nu \alpha \beta} \) is the total
antisymmetric tensor.

According to the factorization, the hadronic tensor contains both short-distance
coefficients and long-distance matrix elements for suitable operators.
Based on NRQCD, the hadronic tensor can be factored as:
\begin{eqnarray}
t_{\mu\nu}(P,k) & \;\equiv\; &
 {\rm Im} \int d^4x e^{ikx}
 \langle B_c(P)|J_\mu^w(x)J_\nu^w(0)
 |B_c(P)\rangle \nonumber \\
 &\;= \;& \Big[\;
 d^{\mu\nu}_1(^1S_0;P,k)\;
  \frac{1}  {M^2}\;
  \langle B_c |
  \psi_c^\dag  \chi_b\; \chi_b^\dag  \psi_c
  | B_c \rangle
  \nonumber \\
  &&  \;+\;
   d^{\mu\nu}_8(^3S_1;P,k)\;
    \frac{1} {M^2}\;
  \langle B_c |
  \psi_c^\dag \sigma^i T^a \chi_b\; \chi_b^\dag \sigma^i T^a \psi_c
  | B_c \rangle
  \nonumber \\
  &&   \;+\;
  d^{\mu\nu}_8(^1P_1;P,k)\;
    \frac{1} {M^4}\;
  \langle B_c |
  \psi_c^\dag  (-i\frac{1}{2}\stackrel
{\leftrightarrow}{D})T^a \chi_b\; \chi_b^\dag
(-i\frac{1}{2}\stackrel {\leftrightarrow}{D}) T^a \psi_c
  | B_c \rangle \;\Big] \;,
 \label{eq:oper}
\end{eqnarray}
here $d^{\mu\nu}_1(^1S_0;P,k)$, $d^{\mu\nu}_8(^3S_1;P,k)$ and
$d^{\mu\nu}_8(^1P_1;P,k)$ are the factors of the short-distance
coefficients which we need to compute out precisely.
Comparing Eq.(\ref{eq:inwidth}) with Eq.(\ref{fact}), they
are related to the coefficients $ C_1(^1S_0), C_8(^3S_1)$
and $ C_8(^1P_1)$ in Eq.(\ref{fact}) precisely as follows:
\begin{eqnarray}
 \label{c1-d1}
 C_1(^1S_0)  &\;=\;&
 \frac{G_F^2V_{cb}^2}{2(2\pi)^6}\int \frac{d^3\vec{k}_4}{{2
 \varepsilon_{4}}}\frac{d^3\vec{k}_5}{{2 \varepsilon_{5}}}
 L_{\mu\nu}(k_4,k_5) \cdot d_1^{\mu\nu}(^1S_0; P,k) \;,
\\
\label{c8-d8}
 C_8(^3S_1)  &\;=\;&
 \frac{G_F^2V_{cb}^2}{2(2\pi)^6}\int \frac{d^3\vec{k}_4}{{2
 \varepsilon_{4}}}\frac{d^3\vec{k}_5}{{2 \varepsilon_{5}}}
 L_{\mu\nu}(k_4,k_5) \cdot d_8^{\mu\nu}(^3S_1; P,k) \;,
\\
\label{c8p-d8p}
 C_8(^1P_1)  &\;=\;&
 \frac{G_F^2V_{cb}^2}{2(2\pi)^6}\int \frac{d^3\vec{k}_4}{{2
 \varepsilon_{4}}}\frac{d^3\vec{k}_5}{{2 \varepsilon_{5}}}
 L_{\mu\nu}(k_4,k_5) \cdot d_8^{\mu\nu}(^1P_1; P,k) \;,
\end{eqnarray}
with $k=k_4+k_5$.

These coefficients can be evaluated by using the threshold
expansion method\cite{chenyq0}. In the method, on-shell $(c\bar{b})$
pair near the threshold with specified quantum number are annihilated.
The amplitudes for the processes at the quark level are calculated
in the fundamental pQCD and expanded as the relative velocity $v$ in
the framework of factorization formula. In the meantime, the real process
of the $B_c$ annihilation may be calculated in the framework of
the effective theory NRQCD and `suffers' a similar factorization,
in which the short distance coefficients are the same. By matching these
two ways of calculations, the short distance coefficients can be
`read off'.

More precisely the coefficients $C_1(^1S_0)$ and
$d_1^{\mu\nu}(^1S_0; P,k)$ are determined by matching
the process $(c\bar{b})_1(^1S_0) \rightarrow l^{+}\nu_{l} \, +gg $,
while $C_8(^3S_1)$ and $d_8^{\mu\nu}(^3S_1; P,k)$;
$ C_8(^1P_1)$ and $ d_8^{\mu\nu}(^1P_1; P,k)$ are determined by
matching the processes $(\bar{b}c)_8(^3S_1) \rightarrow l^{+} \nu_{l} \,
+g$; $(\bar{b}c)_8(^1P_1) \rightarrow l^{+} \nu_{l} \,+g$ with
the calculations on the annihilation of the meson $B_c$ with
the effective theory NRQCD respectively.
In the following two subsections, we show how to determine these
coefficients by matching relevant processes.

\subsection{A. Short distance coefficient for color-singlet}

\begin{figure}
\setlength{\unitlength}{1mm}
\begin{picture}(80,60)(30,30)
\put(-10,-35){\includegraphics{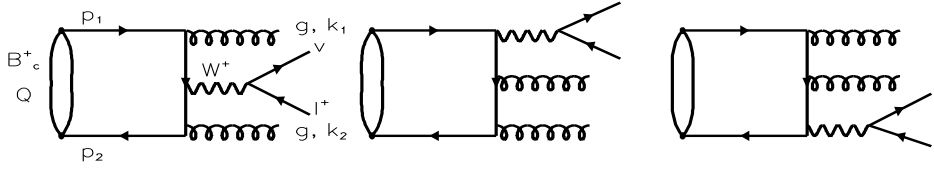}}
\end{picture}
\caption{Three typical Feynman diagrams for the decay
\(B_{c}\rightarrow l^{+}\nu g g\).} \label{singfey}
\end{figure}

In this subsection, we use the threshold expansion method to
determine the short distance coefficients  $C_1(^1S_0)$ and
$d_1^{\mu\nu}(^1S_0; P,k)$ by matching the process $c(p_1)
\bar{b}(p_2) \rightarrow l^{+} (k_4) \nu_{l} (k_5) \, +g (k_1)
g(k_2)$, where $(c\bar{b})$ is in a color-singlet state. At the
lowest order, there are only six Feynman diagrams contributing to
the matrix element. Three of them are shown in Fig.\ref{singfey}
and the other threes can be obtained by exchanging the gluons. The
amplitude is given by:
\begin{eqnarray}
  M &\;=\;&\frac{G_{F}V_{cb}}{\sqrt{2}} 
  \langle l\nu_{l} | j_{\mu} | 0  \rangle
  \; A^\mu\;,
\end{eqnarray}
where $A^\mu$ is defined by
\begin{eqnarray}
 A_\mu &\;\equiv \;&
 \langle gg | J_{\mu} | (\bar{b}c)_1(^1S_0) \rangle \;=\;
  \sum_{i=1,2,\cdot,6} A_{\mu,i}
\end{eqnarray}
The six terms in the amplitude correspond to six Feynman diagrams,
respectively. The first three, which correspond to the diagrams
in Fig.\ref{singfey}, are given by:
\begin{eqnarray}
  A_{\mu,1} &\;=\;& \frac{g_{s}^2\delta^{ab}}{2\sqrt 3} \;\bar{b} (p_2) \;
         \slash\!\!\! \varepsilon^b(k_2)  \frac{1}
        {\slash\!\!\! k_{2}-\slash\!\!\! p_{2}-m_{b}}\gamma_{\mu} (1-\gamma_{5})
        \frac{1}{\slash\!\!\! p_{1}-\slash\!\!\! k_{1}-m_{c}}
         \slash\!\!\!\varepsilon^a(k_1)  \;c(p_1)\;,
        \label{eq:trace1} \\
  A_{\mu,2}&\;=\;& \frac{g_{s}^2\delta^{ab}}{2\sqrt 3} \;\bar{b} (p_2)\;
         \slash\!\!\!\varepsilon^a(k_1)  \frac{1}
        {\slash\!\!\! p_{2}-\slash\!\!\! k_{1}-m_{b}}
        \slash\!\!\! \varepsilon^b(k_2)
        \frac{1}{\slash\!\!\! p_{1}
        -\slash\!\!\! k+m_{b}}\gamma_{\mu} (1-\gamma_{5})\;
        c(p_1) \;,
        \label{eq:trace2} \\
  A_{\mu,3} &\;=\; & \frac{g_{s}^2\delta^{ab}}{2\sqrt 3} \;\bar{b} (p_2) \;
   \gamma_{\mu} (1-\gamma_{5}) \frac{1}
        {\slash\!\!\! p_{2}-\slash\!\!\! k+m_{c}}
        \slash\!\!\! \varepsilon^b(k_2)
        \frac{1}{\slash\!\!\! p_{1}-\slash\!\!\! k_{1}-m_{c}}
        \slash\!\!\!\varepsilon^a(k_1)  \;c(p_1) \;,
        \label{eq:trace3}
\end{eqnarray}
where $\bar{b}(p_2)$, $c(p_1)$ are the Dirac four-component spinors
of the antiquark $\bar{b}$ and the quark $c$ respectively.
$\varepsilon^a(k_1)$ and $\varepsilon^b(k_2)$ are the polarization
vectors of the two gluons with color indices $a,b$ respectively.
The other three terms of the amplitude
can be obtained by exchanging the gluon vertices.

The momenta of \( \bar{b} \), \( c \) quarks are related to their
total and relative momenta $P$ and $q$ as follows:
\begin{equation}
   p_{1}= \mu_c P+q\;,~~~
   p_{2}= \mu_b P-q\;
   \label{eq:momentum}
\end{equation}
here $\mu_c \equiv \frac{m_{c}}{m_{c}+m_{b}}$
and $\mu_b \equiv \frac{m_{b}}{m_{c}+m_{b}}$.
In the case of the lowest order calculations for the $S-$wave, that is we are
considering, the relative momentum $q$ can be set to null, i.e. $q=0$,
hence the momenta can be further simplified as
\begin{displaymath}
  p_{1}=\mu_c P\;, ~~~
  p_{2}=\mu_b P\;.
\end{displaymath}

Then the hadronic tensor $t^{\mu\nu}(P,k)$ for the annihilation
$[(c\bar{b})_1(^1S_0)] \to l^++\nu_l+g+g$ can be expressed as
\begin{eqnarray}
 t^{\mu\nu} (P,k) &\;=\;&
 \frac{1}{(2\pi)^2}\int \frac{d^{3} \vec{k}_{1}}
         {2 \varepsilon_{1}} \frac{d^{3} \vec{k}_{2}}{2 \varepsilon_{2}}
          \;A^{\mu*} A^\nu \; \delta^4 (P-k-k_{1}-k_{2})
 \nonumber \\
 & \;=\;& d^{\mu\nu}_1(^1S_0;P,k)\;
  \frac{1}  {M^2}\;
  \langle (c\bar{b})_1(^1S_0) |
  \psi_c^\dag  \chi_b\; \chi_b^\dag  \psi_c
  | (c\bar{b})_1(^1S_0) \rangle + \cdots \;,
\label{t-bc}
\end{eqnarray}
where ``$\cdots$'' denotes those terms, which, as the ones corresponding
to the pure leptonic decays and those corresponding to the $(c\bar{b})$ with
other quantum numbers, we are not interested in this paper.

The integration over the gluon phase space can be preformed. The
details of the calculations are given in the Appendix A, but we
outline the basic steps here. By introducing three
Lorentz invariant variables: \(z=\frac{(k_1+k_2)^{2}}{M^{2}}\),
\(y=\frac{k^{2}}{M^{2}}\) and \( x_{l}=\frac{2k_4\cdot P}{M^2} \),
the hadronic tensor $t^{\mu\nu} (P,k)$ can be expressed in
terms of them. The short distance coefficients
$d^{\mu\nu}_1(^1S_0;P,k)$ can then be obtained by matching the
Eq.(\ref{t-bc}) and Eq.(\ref{eq:oper}).

The most general form of the tensor $d^{\mu\nu}_1(^1S_0;P,k)$ can be
expressed as:
\begin{eqnarray}
 d^{\mu\nu}_1(^1S_0;P,k) &\;=\;& A g^{\mu \nu}+B P^{\mu} P^{\nu}+C
 P^{\mu} k^{\nu}+D k^{\mu} P^{\nu}+E k^{\mu} k^{\nu}\; .
\label{d-mn}
\end{eqnarray}
However, when we contract it with the leptonic tensor Eq.(\ref{l-t}),
the contribution from the last three terms of Eq.(\ref{d-mn}) vanish
due to the fact that we ignore the lepton masses totally, so the leptonic
weak current (even including the axial component) are conserved.
Namely we have $k_\mu L^{\mu\nu}(k_4,k_5)=0$.
Thus only the first two terms are effective. The detail calculations
and the expressions of the coefficients $A$ and $B$ are put in Appendix A.

Since we are interested in the energy spectrum of the charged lepton which
is measurable, so we also give the coefficients before carrying out the integration
over the leptonic spectrum $x_l$, and those before the integration over the $z, y$
and $x_l$:
\begin{equation}
C_1(^1S_0) \;=\; \frac{G_F^2V_{cb}^2M^4}{2^9\pi^4}\int d y d z d x_{l}
(d_1^{\mu \nu}(^1S_0;P,k) L_{\mu \nu}) \vartheta(x_{l})
\vartheta(1+y-z-x_{l}) \vartheta(Y),
\end{equation}
where the new variable
$$Y\equiv \frac{(1+y-z-x_{l})^{2} x_{l}^{2}}{4}-(y-\frac{(1+y-z-x_{l})
x_{l}}{2})^{2}$$
is introduced. By setting \(Y=0\) and drawing the Dalitz diagram for the process,
we may determine the integration area for \(y\), \(z\) and \(E_{l}\)
(or \(x_{l}\)). The result about the integration area is also put in
Appendix. Since the integration is quite complicated
so we carry out it(them) numerically.

If one carry out the integration in the order: to integrate
$y$ and $z$ first, then $x_l$, then, before doing the last integration
for $x_l$, the measurable energy spectrum of the charged lepton
$\frac{d\Gamma}{dx_l}$ can be obtained. Thus we will do the integration
in this order and discuss the obtained energy spectrum in Section IV.

In order to check the numerical results, we also take a different order to
do the integration. Namely we try to make the integration of the variable
of the leptons $y$ at the last, but try to integrate the other variables of
the leptons and those of gluons.  First of all, we need to change
the lepton tensor from \(L^{\mu\nu}\) to \(N^{\mu \nu}\):
\begin{eqnarray}
N^{\mu \nu} &\equiv&
\frac{4 \pi}{3}(k^{\mu} k^{\nu}-k^{2}g^{\mu \nu}) \;.
\end{eqnarray}
To complete the phase space integration in this order, and having all
the other variables, such as angles etc, of the phase space integrated
out, we reach to the step that only three independent integration
variables: \(y\), the sum of the gluons' energies \(\varepsilon_{s}\)
and the difference of the gluons' energies \(\varepsilon_{d}\)
need to be integrated out. Then to carry out the integration further,
we need to determine the integration area for these three variables by
the Dalitz diagram, and the obtained result about the integration area
is \(0 \leq y \leq 1 \), \( -\varepsilon^{max}_{d}\leq \varepsilon_{d}
\leq \varepsilon^{max}_{d} \) and \(-\varepsilon^{min}_{s} \leq \varepsilon_{s}
\leq \varepsilon^{max}_{s} \), where
\begin{eqnarray}
\varepsilon^{max}_{d}&=&\sqrt{(M-\varepsilon_{s})^{2}-M^{2} y}, \nonumber \\
\varepsilon^{max}_{s}&=&M (1-\sqrt{y}), \nonumber \\
\varepsilon^{min}_{s}&=&\frac{M (1-y)}{2}.
\end{eqnarray}
So the coefficient $C_1(^1S_0)$ is written:
\begin{equation}
C_1(^1S_0) \;=\; \int d y
\int_{\varepsilon^{min}_{s}}^{\varepsilon^{max}_{s}}
\int_{-\varepsilon^{max}_{d}}^{\varepsilon^{max}_{d}}
f_{l}(\varepsilon_{s},\varepsilon_{d},y) d \varepsilon_{s} d \varepsilon_{d},
\end{equation}
where the integrand \(f_{l}(\varepsilon_{s},\varepsilon_{d},y)\)
is obtained straightforwardly in this order step by step as
described above. Since the results for each step are very tedious
so we do not present here, but we have done the computations
carefully and really have the check for the phase space integration
numerically. Indeed we find that the final results for the annihilation
rate obtained by numerical integration in these two orders are the same,
so we are sure that the numerical results very well.
In fact, as a semi-finished result in this integration order,
the `spectrum' \(\frac{d\Gamma}{\Gamma_{B_{c}} d y}\) on
\(y=\frac{k^{2}}{M^{2}}\) may be obtained too, but it is not
easy to measure experimentally so we will not present the curves of the
spectrum here.

\subsection{B. Short distance coefficients for color-octet}

\begin{figure}
\centering
\includegraphics[width=5in]{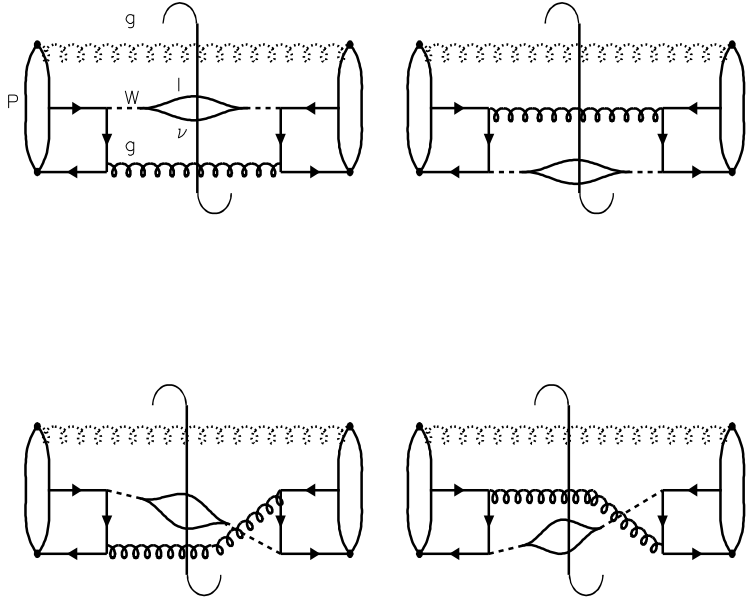}%
\caption{The diagrams for the annihilation width of \(B_{c}\rightarrow l^{+}\nu
g\), where \(B_{c}\) is indicated in a color octet Fock state. The vertical curcul line
in the diagrams  is understood the according imaginary part being taken.}
\label{octetfeyn}
\end{figure}

In this subsection, we use the threshold expansion method to
determine the short distance coefficients  $C_8(^3S_1)$,
$d_8^{\mu\nu}(^3S_1; P,k)$ and $C_8(^1P_1)$, $d_8^{\mu\nu}(^1P_1;P,k)$
by matching the process $c(p_1) \bar{b}(p_2) \rightarrow
l^{+} (k_4) \nu_{l} (k_5) \, +g (k_1)$ where $c\bar{b}$ is in a
color-octet state obviously with the one computed by the effective theory
NRQCD. At the lowest order, there are only two Feynman diagrams
contributing to the amplitude of the process. They are shown
in Fig.\ref{octetfeyn}. The amplitude is given by:
\begin{eqnarray}
M^{a'} &\;=\;&\frac{G_{F}V_{cb}}{\sqrt{2}} 
\langle l\nu_{l} | j^{\mu} | 0  \rangle
\; A^{a'}_{\mu}\;,
\end{eqnarray}
where $A^{a'}_{\mu}$ is defined by
\begin{eqnarray}
A^{a'}_\mu &\;\equiv \;&
\langle g | J'_{\mu} | (\bar{b}c)_8 \rangle \;=\;
\sum_{i=1,2} A^{a'}_{\mu,i}\;,
\end{eqnarray}
where $a=1,\dots8$ are the color indices. The two terms of the amplitude,
corresponding to the two Feynman diagrams shown in
Fig.\ref{octetfeyn}, are given by:
\begin{eqnarray}
  A^{a'}_{\mu,1} &\;=\;& \frac{1}{2}Tr[T^aT^b]{g_{s} } \;\bar{b} (p_2) \;
         \gamma_{\mu} (1-\gamma_{5})
        \frac{1}{\slash\!\!\! p_{1}-\slash\!\!\! k_{1}-m_{c}}
         \slash\!\!\!\varepsilon^b(k_1)\;c(p_1)\;,
        \label{eq:8trace1} \\
  A^{a'}_{\mu,2}&\;=\;& \frac{1}{2}Tr[T^aT^b]{g_{s}} \;\bar{b} (p_2)\;
         \slash\!\!\!\varepsilon^b(k_1)  \frac{1}
        {\slash\!\!\! p_{2}-\slash\!\!\! k_{1}-m_{b}}
       \gamma_{\mu} (1-\gamma_{5})\;
        c(p_1) \;.
        \label{eq:8trace2}
\end{eqnarray}
In the cases of the $S-$wave and $P-$wave $c\bar{b}$ that we are considering
and to the lowest order approximation, we expand the expression in power of
$q$, the relative momentum, and keep the terms only up-to linear ones of $q$
(because we are doing the `leading order calculations only).

Then the hadronic tensor $t^{\mu\nu}(P,k)$ for the annihilation
$(c\bar{b})_1(^1S_0) \to l^+\nu_lg$ can be expressed as
\begin{eqnarray}
t^{\mu\nu} (P,k) &\;=\;&
 \int \frac{d^{3} \vec{k}_{1}}
         {2 \varepsilon_{1}}
          \;A^{a'\mu*} A^{a'\nu} \; \delta^4 (P-k-k_{1})
 \nonumber \\
  &\;=\;&
   d^{\mu\nu}_8(^3S_1;P,k)\;
    \frac{1} {M^2}\;
  \langle (c\bar{b})_8 ^3S_1 |
  \psi_c^\dag \sigma^i T^a \chi_b\; \chi_b^\dag \sigma^i T^a \psi_c
  | (c\bar{b}_8 ^3S_1 \rangle
  \nonumber \\
  &\;+\;&
  d^{\mu\nu}_8(^1P_1;P,k)\;
    \frac{1} {M^4}\;
  \langle (c\bar{b}_8 ^1P_1 |
  \psi_c^\dag  (-i\frac{1}{2}\stackrel
{\leftrightarrow}{D})T^a \chi_b\; \chi_b^\dag
(-i\frac{1}{2}\stackrel {\leftrightarrow}{D}) T^a \psi_c
  | (c\bar{b}_8 ^1P_1 \rangle \;\Big]
  + \cdots \;,
\label{t-bc-8}
\end{eqnarray}
where $\cdots$ denotes the terms corresponding to the $c\bar{b}$
being in the other states, which we are not interested in here.

The integration over the gluon phase space can be performed easily.
Thus with Eq.(\ref{c8-d8}) and Eq.(\ref{c8p-d8p}), the short distance
coefficients can be computed easily.

\begin{equation}
C_8(^{3}S_{1}) = \int d x_l \frac{\alpha_{s} M^6 (m_{b}^{2} +
m_{c}^{2})  G_{F}^2 V_{bc}^2}{24 \pi^2 m_b^2 m_{c}^{2}}
x_{l}\left( x_{l} - 4 (1-x_{l})\log (1 - x_{l}) \right )
\label{eq:8-3S1}
\end{equation}
for the component $((c\bar{b})_8(^{3}S_{1}))$, and
\begin{eqnarray}
C_8(^{1}P_{1}) &=& \int d x_l \frac{\alpha_s M^6 G_{F}^2
V_{bc}^2}{12 \pi^2 m_{b}^4 m_{c}^4 (1-x_{l})}\bigg \{ x_{l}\bigg[
M^2 \big\{-3 m_{b} m_{c}^{3}
(x_{l}-1) + m_{b}^{4}(x_{l}-1)^{2} + m_{c}^{4} (x_{l} -1)^{2} \nonumber \\
&-&m_{b}^3 m_{c}(x_{l} -1)(8x_{l}-7) + 2 m_{b}^{2} m_{c}^{2 }[ 3 +
2(x_{l}-2)x_{l}]\big\} \nonumber\\
&+& 2 M m_{b} m_{c}\big \{ -m_{c}^3(x_{l}-1)(2x_{l}
-7)+m_{b}m_{c}^2(x_{l}-1)(1+2x_{l}) \nonumber\\
&+& m_{b}^3[(7-2x_{l})x_{l}-5]+{ m_{b}}^2 m_{c} [ 1 + x_{l} (6x_{l}-11) ] \big \} \nonumber\\
&+& 2m_{b}^2 m_{c}^2\big \{m_{b}^2[11 +(x_{l}-6)x_{l}] + m_{c}^2[(x_{l}-2)x_{l}-1] +
2 m_{b}m_{c}(x_{l}^2-3)\big \}\bigg]  \nonumber \\
&-&(1-x_{l})\ln (1 - x_{l})\bigg[ M^2 \{ m_{b}^3 m_{c} (7-6 x_{l})
+ m_{b}^4 (x_{l}-1) + m_{c}^4(x_{l}-1)\nonumber \\
&+& 2 m_{b}^2m_{c}^2 (2x_{l} -3)+ m_{b} m_{c}^3(2x_{l}-3)\} \nonumber\\
&-& 2 M m_{b} m_{c}\{5 m_{b}^3(x_{l}-1)  + m_{b} m_{c}^2(x_{l}-1) +m_{b}^2
m_{c}(1-7x_{l})+ m_{c}^3(5x_{l} -7)\} \nonumber\\
&+& 2 m_{b}^2 m_{c}^2 \{m_{c}^2(x_{l}+1)-2 m_{b}m_{c}
(x_{l}-3)+m_{b}^2 (5x_{l}-11)\}\bigg] \bigg \}
\label{eq:8-1P!}
\end{eqnarray}
for the component $((c\bar{b})_8(^{1}P_{1}))$.

\section{III. The long distance effects}

Besides those between $b$ and $\bar c$ quark pair in the initial state (the
interactions make the two quarks into a bound state:the meson $B_c$),
there are also multi soft-gluon interactions between the $c$ and $\bar{c}$ pair
when the $\bar{b}$-quark has decayed into $\bar{c}$ and the pair of $c$ and
$\bar{c}$ has not annihilated yet. Since the interactions owes to
multi soft-gluon exchange, so people generally attribute them as to long
distance effects. In this section, we consider this kind of effects in the
annihilation for the color singlet component $B_c\to l^+\nu_lgg$ only, namely
the possible ones, where the interactions make the $c\bar c$ pair to form a
bound state $\eta_c$ with suitable quantum numbers, because they may be sizable
based on naive order counting but for the concerned octet components the effects
cannot be very great.

\begin{figure}
\setlength{\unitlength}{1mm}
\begin{picture}(80,60)(30,30)
\put(0,-35){\includegraphics{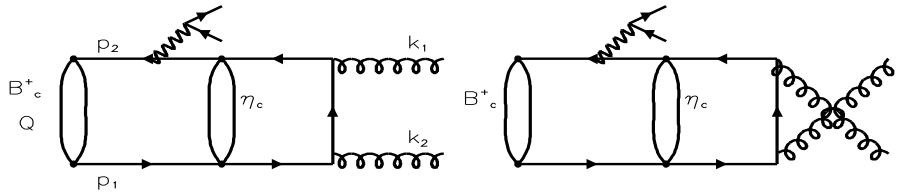}}
\end{picture}
\caption{The diagrams for long distance effects.}
\label{long}
\end{figure}

As a rough estimate, it is enough to consider the dominant effects just to take into account
the possible intermediate meson state \(\eta_{c}\). Now what we are considering is
described in Fig.\ref{long}. The amplitude corresponding to Fig.\ref{long} is
\begin{eqnarray}
M &=& -\frac{G_{F} V_{cb} g_s^{2} }{2 \sqrt{6} }
     (2 \pi)^{4}\delta(P-Q-k)\bar{u}_{\nu}(k_5)
     \gamma^{\mu}(1-\gamma_5) v_{\bar{l}}(k_4)\sum_{\vec{Q}=\vec{k_{1}}
     +\vec{k_{2}}}\frac{\delta_{ab}\epsilon^{a}_{m}(k_{1})
    \epsilon^{b}_{n}(k_{2})}
     {2 Q_{0} (Q_{0}-\omega_{\vec{Q}}+i \epsilon)}  \nonumber \\
     &\cdot &i \int \frac{d^{4}r}{(2\pi)^{4}} \left (Tr \big[ \chi_{Q}(r) \gamma_{m}
   \frac{1}{\frac{1}{2}\slash\!\!\! Q+\slash\!\!\! r-\slash\!\!\! k_{2}-m_{c}}
   \gamma_{n}\big] + Tr \big[ \chi_{Q}(r) \gamma_{n}
   \frac{1}{\frac{1}{2}\slash\!\!\! Q+\slash\!\!\! r-
   \slash\!\!\! k_{1}-m_{c}}\gamma_{m}\big]\right) \nonumber\\
   &\cdot &\int \frac{d^{4}q}{(2\pi)^{4}} Tr \big[\chi_{P}(q) \Gamma_{\mu}
    \bar{\chi}_{Q}(q')(\mu_c \slash\!\!\! P+\slash\!\!\! q +m_{c})\big] \;,
   \label{eq:bound}
\end{eqnarray}
where \(\Gamma_{\mu}=\gamma_{\mu} (1-\gamma_{5})\),
\(\epsilon^{a}_{m}(k_{1})\) and \(\epsilon^{b}_{n}(k_{2})\) are
the polarization vectors of the two real gluons, $q=\mu_bp_1-\mu_cp_2$ and
$q'=q+\frac{1}{2}[(\mu_c-\mu_b)P+k]$ and $r=\frac{1}{2}(p_1-p_2+k)$ are the relative
momenta between the two constitute quarks of the \(B_{c}\) meson and the
bound state $\eta_c$ correspondingly. Note that the equations
$$P=p_1+p_2\;, \;\;\;\;\; k=k_4+k_5\;,\;\;\;\;\; Q=k_1+k_2$$
have been used in Eq.(\ref{eq:bound}).
\(\omega_{\vec{Q}}\equiv \sqrt{\vec{Q}^{2}+(M^{\prime})^{2}}\)
(\(M^{\prime}\) is the mass of $\eta_c$). Note that for the
time-component, usually off mass shell we have
\(Q_{0} \neq \omega_{\vec{Q}}\), but in the present case,
the charged lepton energy may be not very high,
i.e., \(E_{l} \leq \frac{M^{2}-(M^{\prime})^{2}}{2 M} \), so the
intermediate bound state $\eta_c$ may reach to its mass-shell,
that is \(Q_{0} = \omega_{\vec{Q}}\), and leads to a singularity
in Eq.(\ref{eq:bound}). This difficulty happens is due to the fact
that we ignore the width of $\eta_c$. Thus here the width of $\eta_c$
should be considered i.e. to replace $i\epsilon$ with $i\frac{M^{\prime}}{Q_{0}}
\frac{\Gamma_{\eta_{c}}}{2}$ in the relevant propagator, where $i\epsilon$
is the `tiny' imaginary quantity but \(\Gamma_{\eta_{c}}\) is total
width of the bound state \(\eta_{c}\) of $c\bar{c}$.
Then under the non-relativstic approximation, we have
\begin{displaymath}
\frac{1}{2 Q_{0} (Q_{0}-\omega_{\vec{Q}}+i \epsilon)}=
\frac{1}{Q^{2}-(M^{\prime})^{2}+i M^{\prime} \Gamma_{\eta_{c}}}.
\end{displaymath}

\begin{figure}
\setlength{\unitlength}{1mm}
\begin{picture}(60,60)(30,30)
\put(-10,-30){\includegraphics{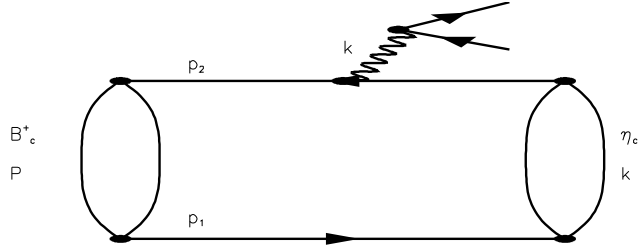}}
\end{picture}
\caption{Current matrix element for the long distance effects.}
\label{matrix}
\end{figure}

Since we consider \(\eta_{c}\) as the only one intermediate bound state
as indicated by Eq.(\ref{eq:bound}), so only the `weak current matrix
element' corresponding to Fig.\ref{long} and the amplitude for
$\eta_c$ annihilation are needed to be computed:
\begin{equation}
\langle \eta_c |\Gamma_{\mu}|P\rangle = i \int \frac{d^{4}q}{(2\pi)^{4}}
 Tr \big[\chi_{B_c}(q) \Gamma_{\mu} \bar{\chi}_{\eta_c}(q')
 (\slash\!\!\! p_{1}+m_{c})\big]\;,
\end{equation}
and
\begin{eqnarray}
 \langle gg|\eta_c \rangle &=& - \frac{ g_s^{2} \delta_{ab}
 \epsilon^{a}_{m}(k_{1})
 \epsilon^{b}_{n}(k_{2})}{2 \sqrt{3}}
 \int \frac{d^{4}r}{(2\pi)^{4}} 
 ( Tr \big[ \chi_{\eta_c}(r) \gamma_{m}
   \frac{1}{\slash\!\!\! p_{1}-\slash\!\!\! k_{2}-m_{c}} \gamma_{n}\big]\nonumber \\
  &+& Tr \big[ \chi_{\eta_c}(r) \gamma_{n}
   \frac{1}{\slash\!\!\! p_{1}-\slash\!\!\! k_{1}-m_{c}} \gamma_{m}\big]
   )\;.
\end{eqnarray}
In fact it is easy to compute the second matrix elements, while to compute
the first one, that corresponds to the Feynman diagram Fig.\ref{matrix},
we need to pay more care to take into account the effects of recoil for
the intermediate state. We adopt the approach, i.e. the so-called generalized
instantaneous approximation, which was proposed firstly in Ref.~\cite{chen},
to deal with the recoil effects in the first current matrix element. The main
points of the approach are to `extend' the potential model, which is based
on Schr\(\ddot{o}\)dinger equation, into the one on Bethe-Salpeter (B.S.)
equation for the non-relativistic binding systems, and then, according to
Mandelstam method~\cite{mandel}, to formulate the current matrix element, so
the current matrix element is set on a full relativistic formulation, finally
by making the `generalized instantaneous approximation' on the whole
relativistic matrix element, i.e., integrating out the `time' component of the
relative momentum in the formulation of the matrix element by a contour
integration, the matrix element turns back to be formulated by means of the
Schr\(\ddot{o}\)dinger wave functions sandwiched by proper operators. The
Schr\(\ddot{o}\)dinger wave functions are just those of the original potential
model for each system and have direct relation to the B.S. wave functions through
the original potential model for each system and have direct relation to the B.S.
wave functions through the original instantaneous approximation proposed by Salpeter.
Since the approach at a middle stage has a fully relativistic formulation for
the weak current matrix (the Mandelstam formulation of the matrix element), so
the final formulation surely takes the recoil effects into account properly.
Since one may find the details of the approximation in several
references\cite{chen,CCWZ,CCL}, and so we do not repeat it here.

By applying the result of the generalized instantaneous approximation
on the current matrix element to computing the matrix elements, one may obtain
the final formula straightforwardly and estimate the long distance effects.

\section{IV. Numerical results and discussions}

We take the parameters as in Refs.\cite{chen,CDF,GEP,ck}:
$\alpha_{s}=0.24$, $\Gamma_{B_{c}}=2.714 ps^{-1}$,
$\psi(0)=0.350 GeV^\frac{3}{2}$,
$\psi^{\prime}(0)=0.250 GeV^\frac{5}{2}$,
in numerical calculations. For comparison, let us list here the
branching ratios of the pure leptonic decays for \(B_{c}\) meson
$$ Br(B_c\to e \nu_e)=1.89\times 10^{-9}\;,\;\;\;
Br(B_c\to \mu \nu_\mu)=7.57\times 10^{-5}\;,$$
$$Br(B_c\to \tau \nu_\tau)=1.95\times 10^{-2}\;.$$
To obtain more reliable values, we should be careful to take the quarks'
masses. According to the discussions in Ref.~\cite{ZL}, here we
take the effective masses of \(c\) and \(b\) quarks to be
\(m_{c}^{eff}=1.5GeV\), \(m_{b}^{eff}=4.9GeV\). For the
mass of \(B_{c}\) meson, we take pole masses of the quarks
\(m_{c}^{pole}=1.88GeV\), \(m_{b}^{pole}=5.02\) ~\cite{ZL,AY,FJY}
first, and then being consistent with that of potential model, the
value of \(B_{c}\) mass \(M=6.352\)GeV is taken.

For the short distance contributions of the process
\(B_{c}\rightarrow ggl\nu\), we find that when \(E_{l}\leq
\frac{m_{b}^{2}-m_{c}^{2}}{2 m_{b}} \), where \(E_{l}\) is the
energy of the charged lepton, being very different from that of
the one photon radiative correction~\cite{CCL}, the \(\bar{c}\)
quark from the decay of \(\bar{b}\) may reach to mass-shell. Thus
to obtain meaningful result, here we should also keep the `width'
of \(c\) quark in its propagator, namely we should make the
following replace:
\begin{displaymath}
 \frac{\slash\!\!\! q +m_{c}}{q^{2}-m_{c}^{2}+i\epsilon} \rightarrow
 \frac{\slash\!\!\! q+m_{c}}{q^{2}-
 m_{c}^{2}+i (4 m_{c} \Gamma_{c})},
\end{displaymath}
where \(\Gamma_{c}\) is the total width of an on-shell $c$-quark and
from $D$ decays its value should be \(\Gamma_{c}\simeq 1.229\)ps$^{-1}$~\cite{MG}.

For the annihilation of color singlet component, \(B_{c}\rightarrow ggl\nu\),
since the color singlet matrix element $\langle B_c |\psi_c^\dag
\chi_b\; \chi_b^\dag  \psi_c | B_c \rangle$ may be related to the wave
function at origin squared $|\psi(0)|^2$
and if only the short distance contributions are taken into account,
the branching ratio can be computed out quite precisely:
\begin{equation}
Br^{short}(B_c\to l\nu_lgg)=2.71 \times 10^{-2}\;.
\end{equation}
Whereas if only the long distance contributions are concerned
as in Sec.III, the branching ratio may be computed out too:
\begin{equation}
Br^{long}(B_c\to l\nu_lgg)=4.45 \times 10^{-3}\;.
\end{equation}
From the above values one may see that the long distance effects are quite great
here. Note that there is slight overlapping for the short distance contributions and
the long distance contributions and these two kinds of contributions should have
interference, but here we ignore all of them.

For the annihilation of color octet components, $B_c \rightarrow l\nu
g$, there is no reliable way to calculate the color octet
matrix elements $\langle B_c |\psi_c^\dag \sigma^i T^a \chi_b\;
\chi_b^\dag \sigma^i T^a \psi_c | B_c \rangle$ and
$\langle B_c | \psi_c^\dag  (-i\frac{1}{2}\stackrel
{\leftrightarrow}{D})T^a \chi_b\; \chi_b^\dag
(-i\frac{1}{2}\stackrel {\leftrightarrow}{D}) T^a \psi_c
| B_c \rangle$ which are necessary when computing the annihilation.

In order to have order estimate roughly, we, based on the velocity
scale rules of NRQCD, try to assume them to be smaller by certain
order $O(v)$ than the S-wave wave functions at the origin
$|\psi(0)|^2$ for color singlet, and the derivative of the P-wave
function at origin $|\psi^{\prime}(0)|^2$, which can be computed
by potential model. Namely based on the velocity scale rule, we
assume
$$\langle B_c |\psi_c^\dag \sigma^i T^a \chi_b\;
\chi_b^\dag \sigma^i T^a \psi_c | B_c \rangle \simeq \Delta_S^2 \cdot
\langle B_c |\psi_c^\dag \sigma^i  \chi_b\; \chi_b^\dag  \sigma^i \psi_c
| B_c \rangle$$
and
$$\langle B_c | \psi_c^\dag  (-i\frac{1}{2}\stackrel
{\leftrightarrow}{D})T^a \chi_b\; \chi_b^\dag
(-i\frac{1}{2}\stackrel {\leftrightarrow}{D}) T^a \psi_c
| B_c \rangle\simeq \Delta_P^2 \cdot \langle B_c |\psi_c^\dag
(-i\frac{1}{2}\stackrel
{\leftrightarrow}{D})\chi_b\; \chi_b^\dag  (-i\frac{1}{2}\stackrel
{\leftrightarrow}{D}) \psi_c | B_c \rangle$$
with $\Delta_S, \Delta_P$ being $O(v)$ constants.
So with the assumption we may evaluate the color octet contributions
accordingly.

In order to explore the characteristics of the color octet components
in the decay for the meson $B_c$ and to have a comparison with those of the
color singlet, we try two possible choices for the S-wave color octet wave
function and the derivative of the P-wave color octet wave function
at the origin: {\bf CASE A}: $\Delta_S\simeq \Delta_P\simeq 0.1$ and {\bf CASE B}:
$\Delta_S\simeq \Delta_P\simeq 0.3$,
because we think that according to the velocity scale rule the range from {\bf CASE A}
to {\bf CASE B} is reasonable and it is helpful to see the possibility if
the color octet components of the meson $B_c$ can be detectable or not
experimentally. The branching ratio of the color octet component annihilations
for {\bf CASE A}:
\begin{equation}
Br(B_c(c\bar{b}_8(^{3}S_{1}))\to l\nu_lg)=1.73 \times 10^{-4}\;,
\;\;\; Br(B_c(c\bar{b}_8(^{1}P_{1}))\to l\nu_lg)=2.24\times
10^{-5}\;\,
\end{equation}
and for {\bf CASE B}:
\begin{equation}
Br(B_c(c\bar{b}_8(^{3}S_{1}))\to l\nu_lg)=1.55 \times 10^{-3}\;,
\;\;\; Br(B_c(c\bar{b}_8(^{1}P_{1}))\to l\nu_lg)=2.02 \times
10^{-4}\;.
\end{equation}

From the values above, we see that the helicity suppression in the
annihilation processes of \(B_{c}\) is released. In total, the
color singlet annihilation to light hadrons is bigger than those
of color octet. Furthermore, when there is one more gluon
bremsstrahlung than the pure leptonic decay, there are more
freedoms in energy and quantum number carried by the gluon(s),
that the intermediate, \(\bar{c}\) quark, even a relevant bound
state such as $\eta_c$ etc (compounded by the produced $\bar
c$-quark and the `original' $c$-quark in the meson $B_c$) may
become on shell, thus the width of the concerned annihilations may
be so great as that of the semi-leptonic decays \(b \rightarrow c
l \nu\) or \(B_{c} \rightarrow \eta_{c} l \nu\) correspondingly.

\begin{figure}
\includegraphics[width=3.1in]{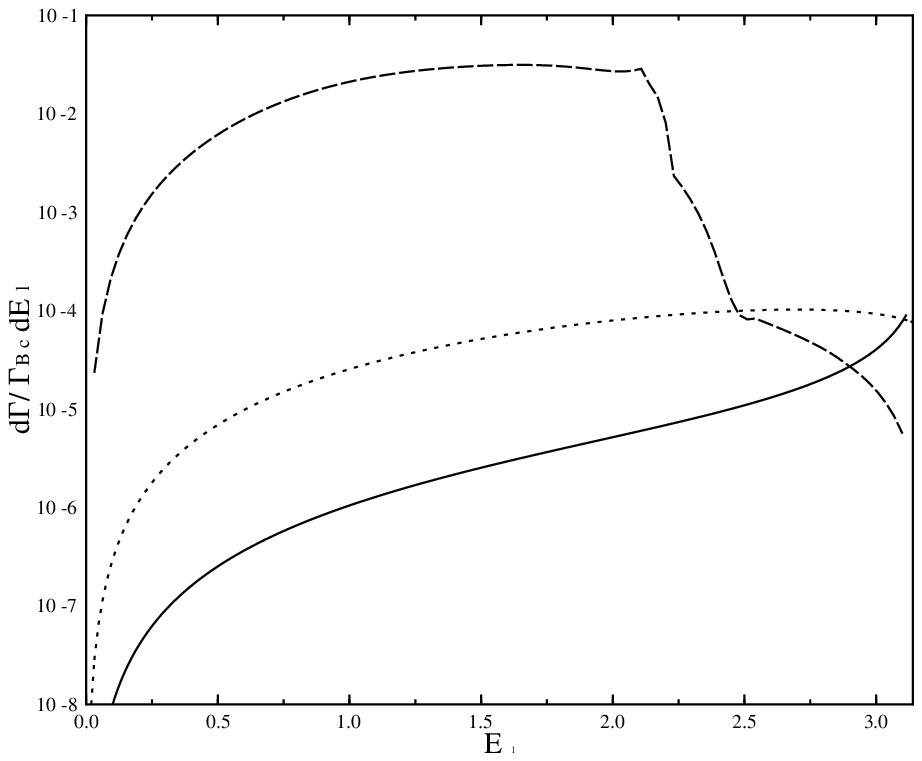}%
\includegraphics[width=3.1in]{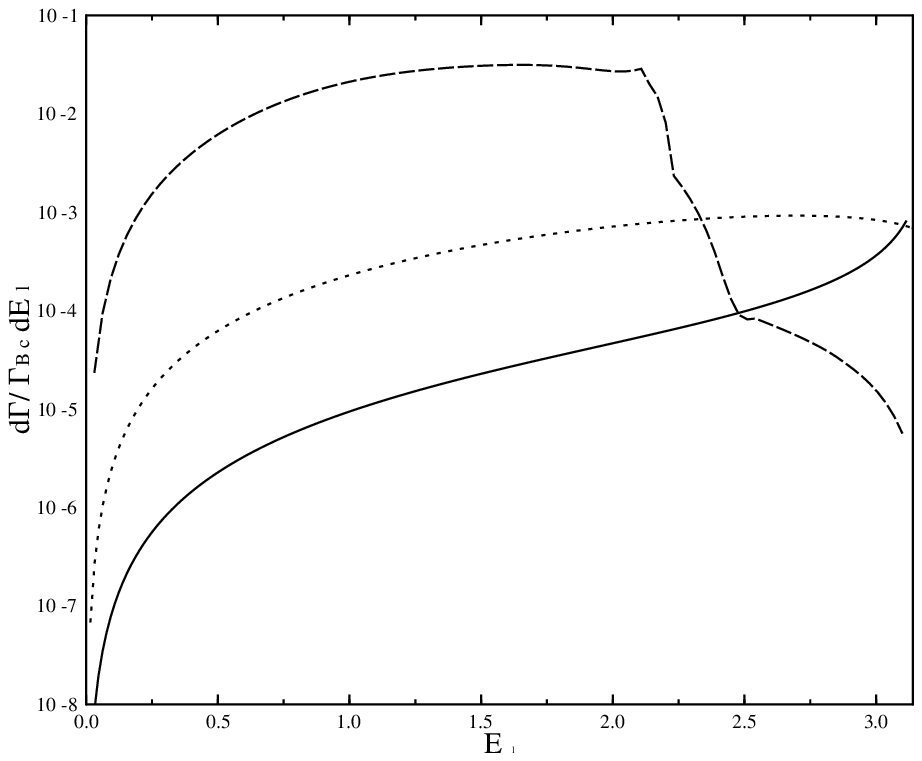}%
\caption{The energy spectra of the charged lepton with different
         `color-octet matrix elements'. The left figure is
         that for {\bf CASE A}: $\Delta_S=0.1$ and $\Delta_P=0.1$.
         The right figure is that for {\bf CASE B}: $\Delta_S=0.3$ and
         $\Delta_P=0.3$. The dashed line in the figures stands for the
         summed effects of short distance contributions and the long
         distance one with \(\eta_{c}\) as an intermediate bound state in
         the color singlet decay \(B_{c} \rightarrow l^{+}\nu g g\).
         The dotted and the solid lines for the color octet annihilations
         \(B_{c} \rightarrow l^{+}\nu g \) with $c\bar{b}_8(^{3}S_{1})$
         and $c\bar{b}_8(^{1}P_{1})$ components in $B_c$ meson
         correspondingly.}
\label{combinelep}
\end{figure}

In Fig.~\ref{combinelep}, we show the lepton energy spectrum of the
annihilation to leptons and light hadrons. The dashed line dictates
that of the spectrum for the color singlet components,
where the short distance and long distance contributions are combined.
The dotted and solid lines, which correspond to the contributions from
the color octet components \((c\bar{b})_8(^{3}S_{1}))\) and
\((c\bar{b})_8(^{1}P_{1}))\) in \(B_{c}\) meson respectively,
dictate the spectra for the color octet components.
It is interesting to point out from Fig.~\ref{combinelep}
that because the meson $B_c$ the wave function of the
color octet components are suppressed, the concerned
annihilations due to the color octet components (one-gluon bremsstrahlung)
even though are smaller comparatively in the most region than that due to the
color singlet component (two-gluon bremsstrahlung), it may become
greater in certain region of the spectrum, so that it is possible
to see the color octet contributions experimentally though studying
the changed lepton energy spectrum of the inclusive decay $B_c\to l^+
\nu_l\cdots$ carefully, especially, around the end point of spectrum.

In order to describe the difference quantitatively in the spectrum
of the charged lepton near the end point, where the color octet
contributions may become dominant over those of the color singlet
ones, we introduce the ratios of the integrated partial decay widths of
$B_{c}$ meson for the color-singlet and the color-octet,
$$R_S=\frac{\Gamma(c\bar{b}_1(^1S_0),x_l^{cut})}{\Gamma(c\bar{b}_1(^1S_0),x_l^{cut})+
\Gamma(c\bar{b}_8(^3S_1),x_l^{cut})}$$ for the S-wave color-octet
component $|c\bar{b}_8(^3S_1)\rangle$, and
$$R_P=\frac{\Gamma(c\bar{b}_1(^1S_0),x_l^{cut})}{\Gamma(c\bar{b}_1(^1S_0),x_l^{cut})+
\Gamma(c\bar{b}_8(^1P_1),x_l^{cut})}$$
for P-wave color-octet state $|c\bar{b}_8(^1P_1)\rangle$, which depend on
the cut of the lepton energy \(x_{l}^{cut}\). Here
$$\Gamma(c\bar{b}_1(^1S_0),x_l^{cut})
\equiv \int_{x_l^{cut}}^{1-\delta}dx\frac{d\Gamma(c\bar{b}_1(^1S_0))}{dx}\;,$$
$$\Gamma(c\bar{b}_8(^3S_1),x_l^{cut})
\equiv
\int_{x_l^{cut}}^{1-\delta}dx\frac{d\Gamma(c\bar{b}_8(^3S_1))}{dx}$$
and
$$\Gamma(c\bar{b}_8(^1P_1),x_l^{cut})
\equiv
\int_{x_l^{cut}}^{1-\delta}dx\frac{d\Gamma(c\bar{b}_8(^1P_1))}{dx}$$
with $\delta=\frac{m_g}{M}$ and a giving tiny gluon mass
$m_g=0.2GeV$. We evaluate them and put result in the Table I.

From Fig.~\ref{combinelep} and Table I, one may see clearly that
there is a possibility to verify experimentally if the color octet
components in the meson $B_c$ play some roles in the $B_c$ annihilation
$B_c\rightarrow l^+\nu_l+${\it hadrons}. It is known that NRQCD is a very
absorbing theory, whereas it needs to verify widely thus we thank that our
estimates are very preliminary but may addresses experimentalists' attention
on this subject. It is worth further to study the possibility in the hadronic
collision environments more quantitatively and the first is Monte Carlo
simulation and then carrying on experimental analysis for the verification.

\vspace*{0.6cm}
\noindent
{\Large\bf Acknowledgements}

\vspace*{0.6cm} \noindent

This work was supported in part by the National Natural Science
Foundation of China (NSFC).

\section{Appendix: Integration formula of the color singlet \(B_{c}\) meson}

A number of kinematic variables will appear repeatedly in the
discussion. For clarity all of them will be collected first.
\(P\), \(k_{1}\), \(k_{2}\), \(k_{4}\), \(k_{5}\) denote the
four-momenta of the various particles, \(Q=k_{1}+k_{2}\), and
\(k=k_{4}+k_{5}\) characterize the gluon-gluon system and the
virtual W respectively.

The scaled masses and lepton energies
\begin{equation}
   y=\frac{k^{2}}{M^{2}}\;,\;\;\; z=\frac{Q^{2}}{M^{2}}\;,\;\;\;
   x_{l}=\frac{2 E_{l}}{M}\;,\;\;\; x_{\nu}=\frac{2 E_{\nu}}{M}
\end{equation}
vary in the region
\begin{eqnarray}
0 \leq & x_{l} & \leq 1\;, \nonumber \\
0 \leq & y & \leq x_{l}\;, \nonumber \\
0 \leq & z & \leq z_{max}=(1-x_{l})(1-\frac{y}{x_{l}})\;,
\end{eqnarray}
where the leptons' mass is ignored. Frequently used kinematical
variables which characterize the gluon-gluon system are
\begin{eqnarray}
 R_{0}&=&\frac{(1 - y + z)}{2}\;, \nonumber\\
 R_{3}&=&\frac{\sqrt{(1-y+z)^{2}-4 z}}{2}\;,\nonumber\\
 Y_{P}&=&\ln \left(\frac{(R_{0}+R_{3})^{2}}{z}\right)\;.
\end{eqnarray}
All of the variables are chosen in such a scaled form on the $B_c$-meson's
mass so as to make no explicit mass dimension left for convenience in the
following defined coefficients \(c_{i}\).

To calculate the annihilation, one observes that the squared
matrix element can be `factorized' into the leptonic tensor
\(L_{\mu \nu}(k_{4},k_{5})\) and the hadronic tensor \(T_{\mu
\nu}(P,k_1,k_2)\equiv A_\mu^*\cdot A_\nu\), ($A_\mu$ is the
amplitude). To perform integration over the gluon-gluon space, the
resultant integration \(t_{\mu \nu}\) depends on \(P\) and
\(Q=k_{1}+k_{2}\) only:
\begin{eqnarray}
t_{\mu\nu}(P,Q)&=&
\int dR_{2}(Q;k_{1},k_{2})T_{\mu
\nu}(P,k_{1},k_{2})\;, \nonumber \\
&=&d_{1,\mu\nu}(^1S_0;P,k)\;
\frac{1} {M^2}\;
\langle B_c |
\psi_c^\dag  \chi_b\; \chi_b^\dag  \psi_c
| B_c \rangle
\end{eqnarray}
where the phase space \(dR_{2}\) is defined by:
\begin{equation}
dR_{2}(Q;k_{1},k_{2})=(2\pi)^4\delta(Q-k_{1}-k_{2})
\frac{d^{3}\vec{k}_{1}}{2(2\pi)^3 E_{1}}\frac{d^{3}\vec{k}_{2}}{2(2\pi)^3 E_{2}}\;,
\end{equation}
and the tensor \(d_{1,\mu\nu}(^1S_0;P,k)\) has the general structure:
\begin{equation}
d_{1,\mu\nu}(^1S_0;P,k)=A g_{\mu \nu}+B P_{\mu} P_{\nu}+C Q_{\mu} Q_{\nu}+D
P_{\mu} Q_{\nu}+E Q_{\mu} P_{\nu}\;.
\label{eq:ttt}
\end{equation}
For the massless lepton pair using the lepton pair current
conservation relation, \(k_{\mu}L^{\mu\nu}=k_{\nu}L^{\mu\nu}=0\)
where \(k=P-Q\) is the total momentum of the lepton pair. With
these equations the above equation Eq.(\ref{eq:ttt}) can be further
simplified as
\begin{equation}
d_{1,\mu\nu}(^1S_0;P,k)=A g_{\mu \nu}+(B+C+D+E) P_{\mu} P_{\nu}=A g_{\mu
\nu}+B^{\prime} P_{\mu} P_{\nu}.
\end{equation}

Instead of integrating the tensor \(T_{\mu \nu}\), it is sufficient
to integrate the following scalar projections
\begin{eqnarray}
c_{1}&=&\frac{M^2}{\langle B_c |
\psi_c^\dag  \chi_b\; \chi_b^\dag  \psi_c
| B_c \rangle}\int dR_{2}T_{\mu \nu}P^{2}g^{\mu \nu}\;, \nonumber \\
c_{2}&=&\frac{M^2}{\langle B_c |
\psi_c^\dag  \chi_b\; \chi_b^\dag  \psi_c
| B_c \rangle}\int dR_{2}T_{\mu \nu}P^{\mu}P^{\nu},
\end{eqnarray}
and the coefficients \(A\), \(B^{\prime}, \cdots\) can be expressed
in terms of the scalar functions \(c_{i}, i=1,2\) as follows:
\begin{equation}
A=\frac{c_{1}-c_{2}}{3P^{2}},\,\,\,
B^{\prime}=\frac{4c_{2}-c_{1}}{3P^{4}}.
\end{equation}

The numerators of \(c_{i}\) are given by polynomials in
\((Pk_{1})\) and \((Pk_{2})\) with coefficients depending on
\(y\), \(z\), \((QP)=M^{2} R_{0}\) and
\((Qk_{1})=(Qk_{2})=\frac{M^{2} z}{2}\). In fact, all of the phase space
integrations may be attributed to the following integrations:
\begin{equation}
I_{m,n}=\int dR_{2}(Q;k_{1},k_{2})(Pk_{1})^{m}(Pk_{2})^{n}  \ \
(-2 \leq m,n \leq 0).
\end{equation}
Because of the Lorentz invariance, we evaluate the integral in the
\(Q=k_{1}+k_{2}\) rest system and the results show
\begin{eqnarray}
I_{0,0}  &=& \frac{1}{8\pi},\nonumber \\
I_{-1,0} &=&I_{0,-1}=\frac{Y_{P}}{8\pi M^{2} R_{3}},\nonumber \\
I_{-2,0} &=&I_{0,-2}=\frac{1}{2\pi M^{4} z},\nonumber \\
I_{-2,-1}&=&I_{-1,-2}=\frac{1}{4\pi M^{6}} (\frac{Y_{P}}{R_{3}
           R_{0}^{2}}+\frac{2}{z R_{0}}),\nonumber \\
I_{-1,-1}&=&\frac{Y_{P}}{4\pi M^{4} R_{0} R_{3}},\nonumber\\
I_{-2,-2}&=&\frac{1}{4\pi M^{8}}
(\frac{Y_{P}}{R_{3}R_{0}^{3}}+\frac{2}{z R_{0}^{2}}).
\end{eqnarray}

The explicit form of \(c_{i}\) are shown in the following, where
a common factor \(\frac{2\pi}{3} \alpha_s^2\) is contracted out
for convenience and the terms that can be obtained by interchanging
\(m_{b}\) and \(m_{c}\) is not shown explicitly, that is, the actual
value of each \(c_{i}\) equals \((c_{i}+c_{i}(r_{1} \leftrightarrow
r_{2}))\), where \(r_{1}=\frac{m_{b}}{M}\) and
\(r_{2}=\frac{m_{c}}{M}\)
\begin{eqnarray}
c_{1} &=& \frac{16}{f_{1}^{2} f_{2} z r_{1}^{2} r_{2}^{2} R_{0}^{2}}
          ( -8 r_{1}^{3} r_{2} ( 3 f_{2} z + 4 ( f_{2}-f_{1} )  r_{2}^{2} )  R_{0}^{2} +
          f_{1}^{2} f_{2} ( y z + 4 R_{0}^{2} )\nonumber\\
      & &    -4 f_{1} r_{1} r_{2} ( f_{1} f_{2} z +
          2 f_{2} ( f_{1} - z )  R_{0} +( 2 f_{1} f_{2} +
          z - 4 f_{2} z + 4 f_{2} r_{2} )  R_{0}^{2} )  +\nonumber\\
      & & 8 r_{1}^{2} ( f_{2} z ( -f_{1} + z )  R_{0}^{2} +
          f_{2} r_{2} R_{0}^{2}( 3 z + ( 4 f_{1} - 2 z )  R_{0} )  -
          4 f_{2} {r_{2}}^3 R_{0} ( 2 f_{1} + R_{0} + R_{0}^{2} )  +\nonumber\\
      & & r_{2}^{2} ( 2 f_{1}^{2} f_{2} + R_{0}^{2} ( 4 f_{2} - 4 f_{1} f_{2} +
          5 f_{1} z - 2 f_{1} R_{0} + 4 f_{2} R_{0}^{2} )  )  )  )  + \nonumber\\
      & & \frac{8{Y_P}}{f_{1}^{2} f_{2} r_{1}^{2} r_{2}^{2} R_{0}^{3} {R_3}}
          ( 16 f_{2} z r_{1}^{4} R_{0}^{2} +f_{1}^{2} f_{2} z ( y - 2 R_{0}^{2} )+\nonumber\\
      & &   4 f_{1} r_{1} ( -( f_{1} f_{2} z r_{2} )  -
          2 f_{2} ( f_{1}-z)r_{2} R_{0}+r_{2} ( 2 f_{1} f_{2}+z (z-1 - 2 f_{2}) \nonumber\\
      & & +4 f_{2} ( 3 - 2 r_{2} )  r_{2} )  R_{0}^{2} +
          2 ( f_{2} z - 2 z r_{2} + 2 f_{2} r_{2}^{2} )
          R_{0}^{3} + 4 r_{2} R_{0}^{4} )+8 r_{1}^{3} r_{2} R_{0}^{2} ( 3 f_{2} z +\nonumber\\
      & &   4 r_{2} ( f_{1} r_{2} + f_{2} R_{0} )  )  +
          8 r_{1}^{2} ( -2 f_{2} z R_{0}^{2} + \nonumber\\
      & & 4 f_{2} {r_{2}}^4 R_{0}^{2} +
          f_{2} ( -2 f_{1} + z )  R_{0}^{4} -
          4 f_{2} {r_{2}}^3 R_{0} ( 2 f_{1} + R_{0}^{2} )  + \nonumber\\
      & & f_{2} r_{2} R_{0}^{2} (z-2 R_{0} ( 2 - 2 f_{1} + z + 2 R_{0}^{2}))
          +r_{2}^{2} ( 2 f_{1}^{2} f_{2} +\nonumber\\
      & &R_{0}^{2} ( -4 f_{1} f_{2} + 4 f_{1} z +
          3 f_{2} z - 4 ( f_{1} - f_{2} )  R_{0} ( 1 + R_{0} )  )  )  )  )  ,
\end{eqnarray}
\begin{eqnarray}
c_{2} &=& \frac{8}{f_{1}^{2} f_{2} z r_{1}^{2} r_{2}^{2}
          R_{0}^{2}} ( f_{1}^{2} f_{2}
          ( z^2 + 8 r_{1} r_{2} ( -z + 4 r_{1} r_{2} )  )  -\nonumber\\
      & &  4 f_{1} f_{2} r_{1} R_{0}( r_{2} ( 4 f_{1} - z R_{0} )  +
          r_{1} ( z R_{0} + 16 r_{2}^{2} ( -2 r_{2} + R_{0} )  )  ) \nonumber\\
      & & + 2 R_{0}^{2} ( f_{1}^{2} f_{2} z + 2 r_{1} ( f_{1} f_{2} z -
          2 f_{1} r_{2} ( z + 3 f_{2} z + 4 f_{2} r_{2} ) -\nonumber\\
      & &   4 r_{1}^{2} r_{2}^{2} ( f_{2} +
          2 ( -f_{1} + f_{2} ) r_{2} - 2 f_{2} R_{0})+ r_{1} ( f_{2} z^2 +\nonumber\\
      & & 2 r_{2}^{2} ( 5 f_{1} z + 2 f_{2} ( y + z )  +
          2 f_{2} r_{2} ( 1 + 2 r_{2} - 2 R_{0} )-4 f_{1} R_{0})))))+ \nonumber\\
      & & \frac{8 {Y_P}}{f_{1}^{2} f_{2} r_{1}^{2} r_{2}^{2} R_{0}^{3} {R_3}}
          ( f_{1}^{2} f_{2} z^2 +8 f_{1}^{2} f_{2} r_{1} r_{2} (-z + 4 r_{1} r_{2} )  -
          16 f_{1} f_{2} r_{1} r_{2} ( f_{1} - 8 r_{1} r_{2}^{2} )  R_{0} +\nonumber\\
      & &  2 ( f_{1}^{2} f_{2} ( 1 - 2 z )  +
          2 r_{1} ( f_{2} z ( f_{1} + ( -2 - 3 f_{1} + 4 y )  r_{1} )  +
          f_{1} ( 4 f_{1} f_{2} +\nonumber\\
      & & ( -3 + f_{2} + y )  z )  r_{2} - 4 ( 2 f_{1} f_{2} + r_{1} ( 8 f_{1} f_{2} -
          ( 3 f_{1} + f_{2} )  z +    \nonumber\\
      & & 2 f_{1} ( -1 + r_{1} )  r_{1} )  )  r_{2}^{2}
          + 8 f_{1} r_{1}^{2} {r_{2}}^3 + 16 f_{2} r_{1} {r_{2}}^4))R_{0}^{2} -\nonumber\\
      & &  16 r_{1} ( -( f_{1} r_{2}( 1 + f_{2} - y + 2 f_{2} r_{2} )  )  -
          r_{1}^{2} ( f_{2} z + r_{2} ( f_{2} -  \nonumber\\
      & & ( f_{1} - 2 f_{2} )  r_{2} )  )  +
          r_{1} r_{2} ( f_{2} ( -4 f_{1} + y + z )  +\nonumber\\
      & &   r_{2} ( 2 f_{1} + f_{2} +
          ( f_{1} + 6 f_{2} )  r_{2} )  )  ) R_{0}^{3} - 16 r_{1}^{2} r_{2}
          ( 2 f_{2} r_{1} + ( f_{1} - 3 f_{2} )  r_{2} )  R_{0}^{4} )   ,
\end{eqnarray}
where \(f_{1}=-1+y+2 r_{1} R_{0} \), \(f_{2}=-1+y+2 r_{2} R_{0}\).

After contraction of the hadronic tensor with \(L_{\mu
\nu}(k_{4},k_{5})\), replacing \(Q\) by \(P-k\) and substituting
\begin{eqnarray}
(k_{4} k_{5})=\frac{M^{2} y}{2}, (k k_{4})=(k k_{5})=\frac{M^{2}
y}{2},
(P k_{4})=\frac{M^{2} x_{l}}{2}, \nonumber \\
(P k_{5})=\frac{M^{2} x_{\nu}}{2}, (Pk)=\frac{M^{2} (1+y-z)}{2}.
\end{eqnarray}
\begin{equation}
x_{\nu}=1+y-z-x_{l}.
\end{equation}
One is left with the task to integrate the function of \(x_{l}\),
\(y\) and \(z\), which can be done numerically.


\newpage
\begin{center}
\begin{table}
\caption{The ratios of the integrated partial width $R_S$ and $R_P$
for {\bf CASE A} ($\Delta_S=0.1, \Delta_P=0.1$) and {\bf CASE B} ($\Delta_S=0.3, \Delta_P=0.3$)
(The definition about $R_S, R_P, \Delta_S, \Delta_P$ and $x_l^{cut}$ is in text)}.

\vspace{4mm}
\begin{tabular}{||c|c|c|c|c||c|c|c|c||}
\hline
 & \multicolumn{4}{c||}{{\bf CASE A:} ($\Delta_S=0.1$)} &
\multicolumn{4}{c||}{{\bf CASE B:} ($\Delta_S=0.3$)} \\
\hline
\multicolumn{1}{||c|}
{$x_l^{cut}$} & 0.80 & 0.85 & 0.90 & 0.95 & 0.80 & 0.85 & 0.90 &
 \multicolumn{1}{c||}{0.95} \\
\hline
\multicolumn{1}{||c|}
{$R_{S}$}   & 0.257 & 0.191 & 0.122 & 0.047 & 0.037 & 0.026 & 0.015 &
\multicolumn{1}{c||}{0.005} \\
\hline
\hline
\hline
  & \multicolumn{4}{c||}{{\bf CASE A:} ($\Delta_P=0.1$)}
  & \multicolumn{4}{c||} {{\bf CASE B:} ($\Delta_P=0.3$)} \\ \hline
\multicolumn{1}{||c|} {$x_l^{cut}$} & 0.90 & 0.92 & 0.94 & 0.96 & 0.80 & 0.85 &
0.90 & \multicolumn{1}{c||}{0.95} \\
\hline
\multicolumn{1}{||c|} {$R_{P}$}& 0.280 & 0.212  & 0.154 & 0.089 &
0.132 & 0.081 & 0.041 & \multicolumn{1}{c||}{0.014}\\ \hline
\end{tabular}

\end{table}
\end{center}

\newpage


\begin{thebibliography}{s2}

\bibitem{CDF} CDF Collaboraten, F.Abe {\em etal.}, Phys. Rev. {\bf D58}, 112004 (1998).

\bibitem{prod} Chao-Hsi Chang and Yu-Qi Chen, Phys. Rev. D {\bf 46}, 3854 (1992);
Erratum Phys. Rev. {\bf 50}, 6013 (1994); Chao-Hsi Chang and Yu-Qi Chen,
Phys. Rev. D {\bf 48}, 4086 (1993); Chao-Hsi Chang, Yu-Qi Chen, Guo-Ping Han
and Hung-Tao Jiang, Phys. Lett. B {\bf 364}, 78 (1995); Chao-Hsi Chang, Yu-Qi Chen
and R. J. Oakes, Phys. Rev. D {\bf 54}, 4344 (1996). E. Braaten, K. Cheung
and T.C. Yuan, Phys. Rev. D {\bf 48}, 4230 (1993); E. Braaten, K. Cheung and
T.C. Yuan, Phys. Rev. D {\bf 48}, R5049 (1993); A.V. Berezhnoy, V.V. Kiselev,
A.K. Likhoded and A.I. Onishchenko, Yad. Fiz. {\bf 60}, 1866 (1997); K. Kolodziej,
A. Leike and R. R\"uckl, Phys. Lett. B {\bf 355}, 337 (1995).

\bibitem{chen} Chao-Hsi Chang, Yu-Qi Chen, Phys. Rev. {\bf D49},3399 (1994);
Chao-Hsi Chang and Yu-Qi Chen, Commun. Theor. Phys. {\bf  23} (1995) 451.

\bibitem{CCWZ} Chao-Hsi Chang, Yu-Qi Chen, Guo-Li Wang and Hong-Shi Zong,
Phys. Rev. D {\bf 65}, 014017 (2001); Commun. Theor. Phys. {\bf 35}, 395 (2001).

\bibitem{dec} M. Lusignoli and M. Masetti, Z. Phys. C {\bf 51}, 549 (1991); N.
Isgur, D. Scora, B. Grinstein and M. Wise, Phys. Rev. D {\bf 39},
799 (1989); D. Scora and N. Isgur, Phys. Rev. D {\bf 52}, 2783
(1995). D.-S. Du, G.-R. Lu and Y.-D. Yang, Phys. Lett. B {\bf
387}, 187 (1996);  Dongsheng Du, {\em etal.}, Phys. Lett. B {\bf 414}, 130(1997);
A. Abd El-Hady, J.H. Munoz and J.P. Vary; Phys.
Rev. D{\bf 62} 014014 (2000); P. Colangelo and F.De Fazio, Phys.
Rev. D {\bf 61} 034012 (2000). V.V. Kiselev, A.E. Kovalsky and
A.K. Likhoded, Nucl. Phys. B {\bf 585} 353 (2000). M.A. Nobes and
R.M. Woloshyn, J. Phys. G {\bf 26} 1079.

\bibitem{MG} M. Beneke and G. Buchalla, Phys. Rev. D {\bf 53}, 4991 (1996).

\bibitem{ZL} Chao-Hsi Chang, Shao-Long Chen, Tai-Fu Feng and
Xue-Qian Li, Phys. Rev. D {\bf 64},
014003 (2001); Commun. Theor. Phys. {\bf 35}, 51 (2001).

\bibitem{chang} Chao-Hsi Chang, Jian-Ping Cheng and Cai-Dian L\(\ddot{u}\),
Phys. Lett B {\bf 425}, 166 (1998).

\bibitem{CCL} Chao-Hsi Chang, Cai-Dian L\"u, Guo-Li Wang
and Hong-Shi Zong, Phys. Rev. D {\bf 60}, 114013 (1999).

\bibitem{chang1} Chao-Hsi Chang, Anjan
K. Giri, Rukmani Mohanta and Guo-Li Wang, J. Phys. G {\bf 28} 1403, (2002).

\bibitem{bct}
M. Kobayashi, T.-T. Lin and Y. Okada, Prog.
Theor. Phys. {\bf 95}, 361 (1996); C. H. Chen, C. Q. geng,
and C. C. Lih, Phys. rev. {\bf D 56}, 6856 (1997); G.-H. Wu
and J. N. Ng, Phys. Rev. {\bf D 55}, 2806 (1997);
C. Q. Geng and S. K. Lee, Phys. Rev. {\bf D 51}, 99 (1995).

\bibitem{TMA} T.M. Aliev and M. Savci, Phys. Lett. {\bf B434}, 358 (1998);
T.M. Aliev and M. Savci, J. Phys. {\bf G25}, 1205 (1999).

\bibitem{GEP} G.T. Bodwin, E. Braaten, and G.P. Lepage, Phys.Rev.D {\bf 51},1125 (1995);
Erratum, {\em ibid} {\bf 55}, 5853 (1997).

\bibitem{ck} Yu-Qi Chen and Yu-Ping Kuang, Phys. Rev. D {\bf 46},1165 (1992);
E. Eichten and C. Quigg, Phys. Rev. D {\bf 49}, 5845 (1994); A.Abd El-Hady,
J.H. Munoz and J.P. Vary; Phys. Rev. D {\bf 55}, 6780 (1997).

\bibitem{chenyq0} E. Braaten and Yu-Qi Chen, Phys. Rev. D {\bf 55}, 2693 (1997).


\bibitem{mandel} S. Mandelstam, Proc. R. Soc. London {\bf 233}, 248(1955).

\bibitem{JXA} J.L. Cortes, X.Y. Pham, and A. Tounsi, Phys. Rev. D {\bf 25}, 188 (1982).

\bibitem{AY} A. Pineda and F.J. Yudurain, Phys. Rev. D {\bf 58}, 094022(1998).

\bibitem{FJY} F.J. Yudurain, hep-ph/0002237.

\end{thebibliography}
\end{document}